# MONITORING AIR MOISTURE WITH LASER ABSORPTION SPECTROSCOPY

# CONTROLE DE L'HUMIDITE DE L'AIR PAR SPECTROSCOPIE D'ABSORPTION LASER

**N-E. Khélifa, P. Pinot**
LNE- INM : Conservatoire national des arts et métiers
*61, rue du Landy, 93210 ; la plaine Saint Denis*

## Résumé

La détermination de la masse volumique de l'air et l'évaluation de l'adsorption- désorption de la vapeur d'eau par la surface des artefacts restent aujourd'hui les principales sources d'incertitudes lors des pesées d'un étalon de référence en acier inoxydable ou en super alliage. Pour la masse volumique de l'air ambiant, la méthode classique consiste à utiliser la formule dite CIPM-1981/91. L'autre approche consiste à déduire la masse volumique de l'air humide à partir des seuls résultats de pesées faites successivement dans l'air et sous vide primaire d'artefacts de géométries bien adaptées. Dans les deux cas, la distribution et l'état des molécules d'eau à l'intérieur de l'enceinte du comparateur de masses influencent les résultats des pesées. Ce travail a permis de montrer que les instruments utilisés pour mesurer l'hygrométrie sont, soit insuffisamment sensibles par rapport au besoin (hygromètre capacitif pour suivre l'évolution de la fraction molaire de vapeur d'eau dans une enceinte fermée), soit perturbants pour l'environnement (hygromètre à point de rosée). Aussi, nous donnerons quelques observations préliminaires sur le comportement de la vapeur d'eau dans l'enceinte, certes médiocre du point de vue de l'étanchéité au vide, au cours du temps et pour différentes conditions expérimentales.

## 1. Introduction

L'interaction de la vapeur d'eau avec les étalons de masse, introduits à l'intérieur de l'enceinte du comparateur de masses, ainsi qu'avec les parois de cette dernière peut conduire à des erreurs dans la détermination de la masse volumique de l'air et des résultats de pesées dans l'air et sous vide. Pour tenter d'observer cet effet, nous avons utilisé le dispositif qui a servi à mettre en évidence les perturbations induites par un hygromètre à condensation. Il est, certes difficile, de déterminer l'erreur due à l'influence de la vapeur d'eau en termes de variation de masse des étalons mis en comparaison. En effet, celle-ci dépend de plusieurs facteurs : forme, surface effective, état de surface des étalons mis en œuvre, de la distribution de la vapeur d'eau dans le volume d'air et les parois de l'enceinte et des différents gradients existants, …. Pour minimiser cet effet, les coefficients d'adsorption et de désorption de la vapeur d'eau par la surface des étalons de masse et de celle de la paroi intérieure de l'enceinte du comparateur doivent être les plus faibles possibles. Enfin, il faut noté que les résultats des vérifications périodiques entre le prototype international de l'unité de masse et les différentes copies disséminées dans le monde montrent parfois des écarts d'évolution pouvant atteindre environ 50 µg. Evidemment, même s'ils sont affectés différemment, les références de masse subissent les effets des processus physico-chimiques de sorption et de désorption d'espèces apportées principalement par l'environnement ($H_2O$, $CO_2$) et par les contacts avec des manipulateurs. L'autre aspect, concerne le niveau d'incertitude avec lequel sont faites les pesées.

Dans l'air, la correction de masse différentielle, induite par la poussée de l'air, est la plus incertaine. Cette dernière est affectée par l'incertitude sur la détermination de la masse volumique de l'air humide. Pour cela, la formule CIPM- 1981/91 est utilisée [1], soit :

$$\rho_a \{CIPM - 1981/91\} = f(p, T, x_v).$$

Où $p$, $T$ et $x_v$ sont respectivement la pression, la température et la fraction molaire de vapeur d'eau dans l'air humide.

Sous vide, la correction de masse, induite par la désorption, constitue une source d'incertitude supplémentaire pour la détermination de $\rho_a$. Ainsi, la méthode gravimétrique permet de déduire la masse volumique de l'air à partir des résultats de pesées d'artefacts dans l'air et sous vide [2, 3], soit :

$$\rho_a^{artefacts} = (\Delta m_{vide} - \Delta m_{air} + \Delta m_{sorption})/\Delta V$$

Où $\Delta m_{vide}$ et $\Delta m_{air}$ sont respectivement les différences de masse sous vide et dans l'air entre les deux étalons ; $\Delta m_{sorption}$ correspond à la différence de masse induite par les processus d'adsorption et $\Delta V$ à la différence de volume entre les deux artefacts.

La correction de masse engendrée par la sorption-désorption de la vapeur d'eau par la surface des étalons est relativement variable et imprécise. En effet, les différents ordres de grandeurs du coefficient d'adsorption par des surfaces en acier inoxydable publiés ou utilisés sont assez différents mais restent inférieurs à $0,5 \ \mu g \ cm^{-2}$ [4, 5, 6]. Comme cela a été déjà signalé, les valeurs de la masse volumique de l'air, obtenues par les deux méthodes, présentent un écart moyen de l'ordre de :

$$\frac{\rho_a^{artefacts} - \rho_a^{CIPM-1981/91}}{\rho_a^{CIPM-1981/91}} \cong 6,5 \times 10^{-5} \ kg \ m^{-3}$$

L'analyse des différentes composantes des incertitudes associées à chacune des deux méthodes laisse apparaître la possibilité d'un biais lors de la mesure de la température du point de rosée (fraction molaire de vapeur d'eau dans l'air).

## 2. Contrôle de la vapeur d'eau dans une enceinte fermée

La figure 1 montre les différentes parties du nouveau dispositif expérimental mis en place pour tenter d'étudier l'évolution de la sorption et de la désorption de la vapeur d'eau par la surface d'artefacts et des parois de l'enceinte. Le système constitué par la source laser et l'instrumentation associée a été précédemment décrit [7, 8].

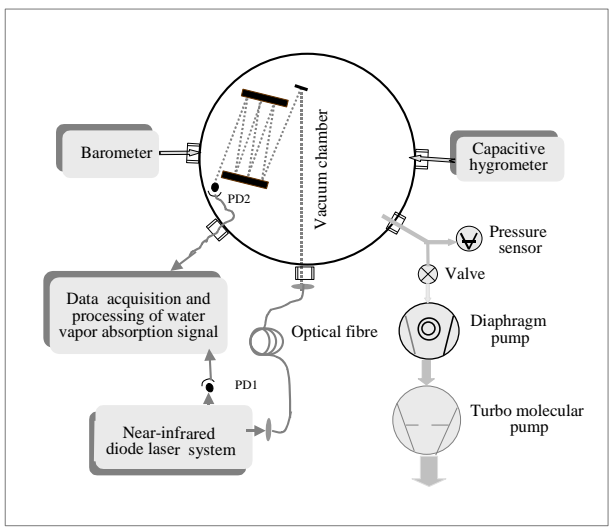

**Figure1.** Schéma de principe du montage expérimental

Le raccordement de l'enceinte au groupe de pompage a été fait par la suite, sur la figure 1, à l'endroit de l'hygromètre à condensation. De ce fait, cet instrument n'apparaît pas sur le schéma du montage expérimental.

Pour valider le dispositif de détection de la vapeur d'eau, basé sur l'absorption moléculaire dans le proche infrarouge ($\lambda \cong 1392,5 \ nm$), nous avons tout d'abord mis en évidence la non transparence de l'hygromètre à point de rosée. La figure 2 donne l'évolution de la vapeur d'eau dans l'enceinte fermée sur plusieurs heures, montrant en particulier les réponses des capteurs à absorption moléculaire et capacitif, lors de la mesure de la température du point de rosée. Ces résultats correspondent aux mesures faites à l'intérieur de l'enceinte contenant de l'air humide caractérisé par les paramètres d'environnement : $p = 101532 \ Pa$, $t = 21,30 \ °C$ et une humidité relative de 52,8 % HR.

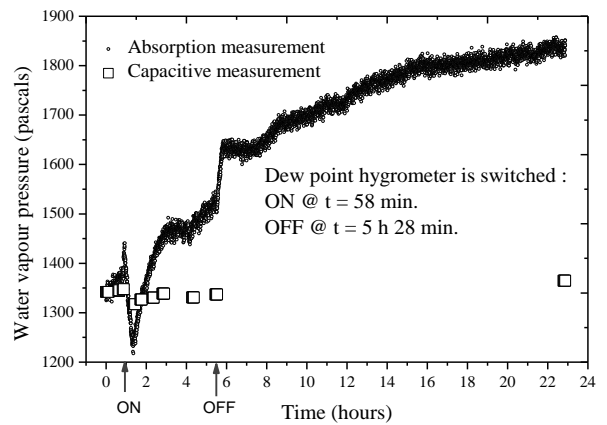

**Figure 2.** Evolution au cours du temps de la vapeur d'eau à l'intérieur d'une enceinte fermée sous pression atmosphérique observée par deux méthodes différentes.

Le signal d'absorption moléculaire mesuré est converti en pression partielle de vapeur d'eau. Pour cela, ce signal est normalisé sur la valeur moyenne de l'humidité relative relevée au début d'une série de mesures. C'est pour cela qu'initialement les deux courbes de la figure 2 sont confondues. La supériorité, du point de vue temps de réponse et sensibilité du capteur par absorption est ainsi confirmée.

## 3. Température du point de rosée et vapeur d'eau dans l'air

La non transparence de l'hygromètre à condensation induit systématiquement une diminution du nombre de molécule d'eau dans l'air sous l'enceinte fermée au cours de la mesure de la température du point de rosée. Aussi, même plusieurs heures après avoir éteint l'hygromètre à point de rosée, les mesures faites avec le transmetteur capacitif montrent un écart par rapport aux observations initiales (avant la mesure de la température du point de rosée). Des résultats de la figure 2, les valeurs de la masse volumique de l'air, obtenues par application de la formule CIPM-1981/91 et de la fraction molaire de vapeur d'eau déduite des mesures par absorption moléculaire et capacitive, peuvent être évaluées pour deux les situations :

i) Avant la mise en route de l'hygromètre à condensation, soit : $\rho_a = 1{,}19561$ kg m$^{-3}$

ii) Bien après avoir allumé puis éteint ce dernier instrument pour la mesure de la température du point de rosée, soit : $\rho_a^{apparent} = 1{,}19566$ kg m$^{-3}$

Ainsi, tout se passe comme si on déterminait une masse volumique de l'air apparent dans le cas où la fraction molaire de vapeur d'eau dans l'air est déduite après la mesure de la température du point de rosée. Cette dernière méthode est, pour le moment, utilisée par l'ensemble des laboratoires nationaux de métrologie. Dans les conditions d'environnement utilisé, un ordre de grandeur de cet écart relatif entre les valeurs de la masse volumique de l'air est :

$$\left\{\frac{\rho_a^{apparent} - \rho_a}{\rho_a^{apparent}}\right\}_{CIPM-1981/91} = 4{,}2 \times 10^{-5}$$

Où $\rho_a^{apparent}$ et $\rho_a$ sont les valeurs de la masse volumique de l'air déterminées à partir de la formule CIPM-1981/91, respectivement avec et sans utilisation de l'hygromètre à point de rosée. Cette situation correspond en quelque sorte à un air apparent plus lourd que l'air qui n'a pas été perturbé par l'hygromètre à condensation.

## 4. Pression partielle de vapeur d'eau dans l'air résiduel

Un système de pompes à vide, pompe sèche associée à une pompe turbo moléculaire est utilisé pour faire varier la pression totale dans l'enceinte de $10^5$ Pa à environ 1 Pa. La pression dans l'enceinte, lorsque le groupe de pompage est arrêté, passe de 1 Pa à 1800 Pa en deux heures. Ici, les processus de sorption et de désorption en fonction du vide dans l'enceinte sont masqués par la fuite globale. Avec l'hygromètre à absorption moléculaire, la quantité de vapeur d'eau dans l'enceinte est enregistrée quand l'air est évacué progressivement de l'enceinte. La figure 3 montre comment évolue la pression partielle de vapeur d'eau dans l'enceinte passant d'une pression de l'ordre de 600 Pa à environ 87130 Pa. L'enceinte est vidée d'air pour être isolée du groupe de pompage. On remarque une sensible diminution de la pression de vapeur d'eau, induite par la combinaison des processus de sorption-désorption des parois de l'enceinte.

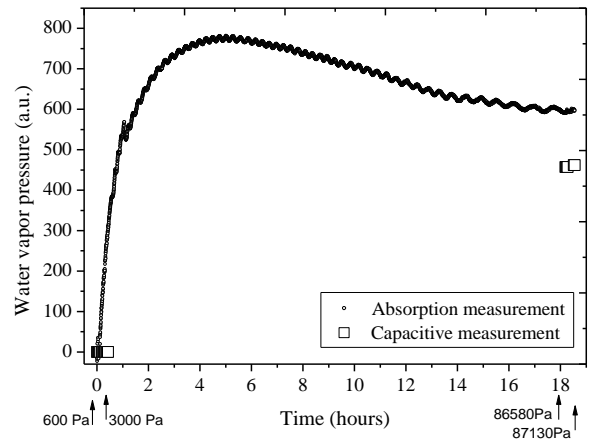

**Figure 3.** Evolution au cours du temps de la vapeur d'eau à l'intérieur d'une enceinte fermée initialement sous vide primaire.

L'air dans l'enceinte est évacué de façon progressive jusqu'à la limite d'utilisation du baromètre, soit environ 87000 Pa. L'enceinte est alors isolée du groupe de pompage. La figure 4 montre l'évolution de la pression partielle de vapeur d'eau dans l'enceinte sur plus de 30 heures, relevée par les deux capteurs. Les résultats montrent que la vapeur d'eau est faiblement affectée, en dessous d'une pression totale d'air de l'ordre de $8 \times 10^4$ Pa,

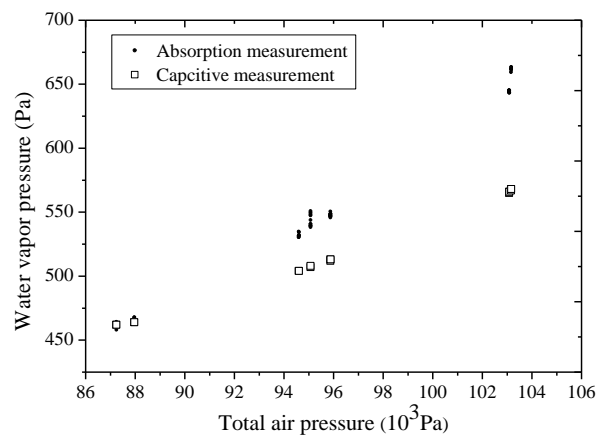

**Figure 4.** Evolution de la pression résiduelle de vapeur d'eau en fonction de la pression totale dans l'enceinte

## 5. Conclusion

Au sens du vocabulaire international des termes fondamentaux et généraux de métrologie (VIM 1993, 3, 16), l'hygromètre à point de rosée n'est pas un instrument transparent. De ce fait, la fraction molaire de vapeur d'eau dans l'air est altérée au cours de la mesure de la température du point de rosée. Ceci peut entraîner une inhomogénéité de l'air sous l'enceinte du comparateur, conduisant à une différence appréciable de l'adsorption de la vapeur d'eau par les étalons et les parois de la chambre. Ce type de dispositif sera implanté sur le comparateur "M One" de portée 1 kg et de résolution 0,1 µg du Laboratoire National de Métrologie et d'Essais (LNE). Les caractéristiques de ce comparateur (meilleure étanchéité au vide de l'enceinte, possibilité de faire des pesées avec une très bonne résolution) permettront d'affiner les résultats de cette étude. L'objectif final est d'élucider les écarts observés entre les valeurs de la masse volumique de l'air obtenues par les deux méthodes déjà citées.

## Références